\documentclass[twocolumn,showpacs,amsmath,amssymb,aps,prl,superscriptaddress]{revtex4-1}

\usepackage[]{cleveref}
\usepackage[]{graphicx}
\usepackage[]{verbatim}
\usepackage[]{color}
\usepackage[abs]{overpic}
\usepackage{rotating}
\usepackage{tabularx}

\usepackage{amsfonts}
\usepackage{amsmath}
\usepackage{amssymb}
\usepackage{bm}
\usepackage{graphicx}
\usepackage[breaklinks,colorlinks=true]{hyperref}
\usepackage{mathrsfs}
\usepackage[lofdepth,lotdepth,caption=false]{subfig}
\usepackage{varwidth}
\usepackage{wrapfig}
\usepackage{times}
\usepackage{longtable}
\usepackage{multirow}
\usepackage{tikz}
\definecolor{darkblue}{RGB}{0 60 120}
\definecolor{eggplant}{RGB}{190 10 150}
\definecolor{darkgray}{RGB}{70 70 70}
\definecolor{lightgray}{RGB}{80 80 80}
\definecolor{lightgray2}{RGB}{245 215 110}
\definecolor{lightgray3}{RGB}{255 0 0}

\graphicspath{{../}}

\begin{document}

\title{Molecular orbital polarization in Na$_2$Ti$_2$Sb$_2$O: \\
microscopic route to metal-metal transition without spontaneous symmetry breaking}

\author{Heung-Sik Kim}
\affiliation{Department of Physics and Center for Quantum Materials , University of Toronto, 60 St.~George St., Toronto, Ontario, M5S 1A7, Canada}

\author{Hae-Young Kee}
\affiliation{Department of Physics and Center for Quantum Materials , University of Toronto, 60 St.~George St., Toronto, Ontario, M5S 1A7, Canada}
\affiliation{Canadian Institute for Advanced Research / Quantum Materials Program, Toronto, Ontario MSG 1Z8, Canada}

\begin{abstract}
Ordered phases such as charge- and spin-density wave state
accompany either full or partial gapping of Fermi surface (FS) leading a metal-insulator or metal-metal
transition (MMT). However, there are examples of MMT without any signatures of symmetry breaking.
One example is Na$_2$Ti$_2$Sb$_2$O, where a partial gapping of FS is observed but a density wave ordering 
has not been found. Here we propose a microscopic mechanism of such a MMT which occurs due to
a momentum dependent spin-orbit coupled molecular orbital polarization.
Since a molecular $d$ orbital polarization is present due to a small spin-orbit coupling of Ti, there is no
spontaneous symmetry breaking involved. However, a sharp increase of polarization happens above a critical
electron interaction which gaps out the $d$ orbtial FS and reduces the density of states significantly, 
while the rest of FS associated with Sb $p$ orbtials is almost intact across MMT. 
Experimental implications to test our proposal and applications to other systems are also discussed.
\end{abstract}


\maketitle

\textit{Introduction} -- 
Fermi surface (FS) is one of most important concepts in solid state physics,
and exists in every metal, semimetal, and doped semiconductors. Many ordered phases
such as charge- and spin-density waves, and superconductivity can be regarded
as FS instabilities. Such ordered phases accompany broken symmetries
involving either translation, time-reversal, or charge conservation symmetries. 
As a consequence, either full or partial gapping of FS occurs with anomalies
in spin susceptibility, speicific heats, and resistivity at a critial 
temperature $T_c$\cite{DW}.

However, there are examples of metallic systems which undergo a phase transition to a metallic state
with a partial gapping of FS at lower temperatures without any detection of spontaneous symmetry breaking. 
A widely studied example is URu$_2$Si$_2$
where both $f$ and $d$ orbitals are relevant and a low-temperature metallic phase is associated 
with a hidden order, implying difficulties of identifying the order parameter\cite{URuSi_first,RMP_URuSi}.
A less studied material is Na$_2$Ti$_2$Sb$_2$O (NTSO), where Ti $d$ orbitals and Sb $p$ orbtials play 
a major role in determining physical properties\cite{NTSO_first,NTSO_second}. 
NTSO shows a metal-metal transition (MMT) around $T_c \simeq 115$ K, where the temperature dependence of
the resistivity above and below $T_c$ indicates its metallic behavior,
while the increase of resistivity and reduction of susceptibility at $T_c$ imply the reduction of
density of states (DOS) at the Fermi level\cite{Ozawa2000,Liu2009,Huang2013}. 
Specific heat versus $T$ shows a sharp peak at $T_c$, signifying substantial 
entropy change\cite{Shi2013}. Previous studies have proposed 
possible charge/spin-density wave instabilities for the origin of 
MMT\cite{Pickett1998,Singh2012,Yan_SDW}, 
however there is no experimental evidence of charge/spin-density wave order.
$^{23}$Na nuclear magnetic resonance measurement revealed no sign of enhanced spin flcutations
or magnetic order\cite{Fan_NMR}, and neutron and x-ray diffraction data found only changes in 
lattice constants across $T_c$\cite{Ozawa2000,Ozawa2004}.



\begin{figure}[htb!]
  \centering
  \includegraphics[width=0.47 \textwidth]{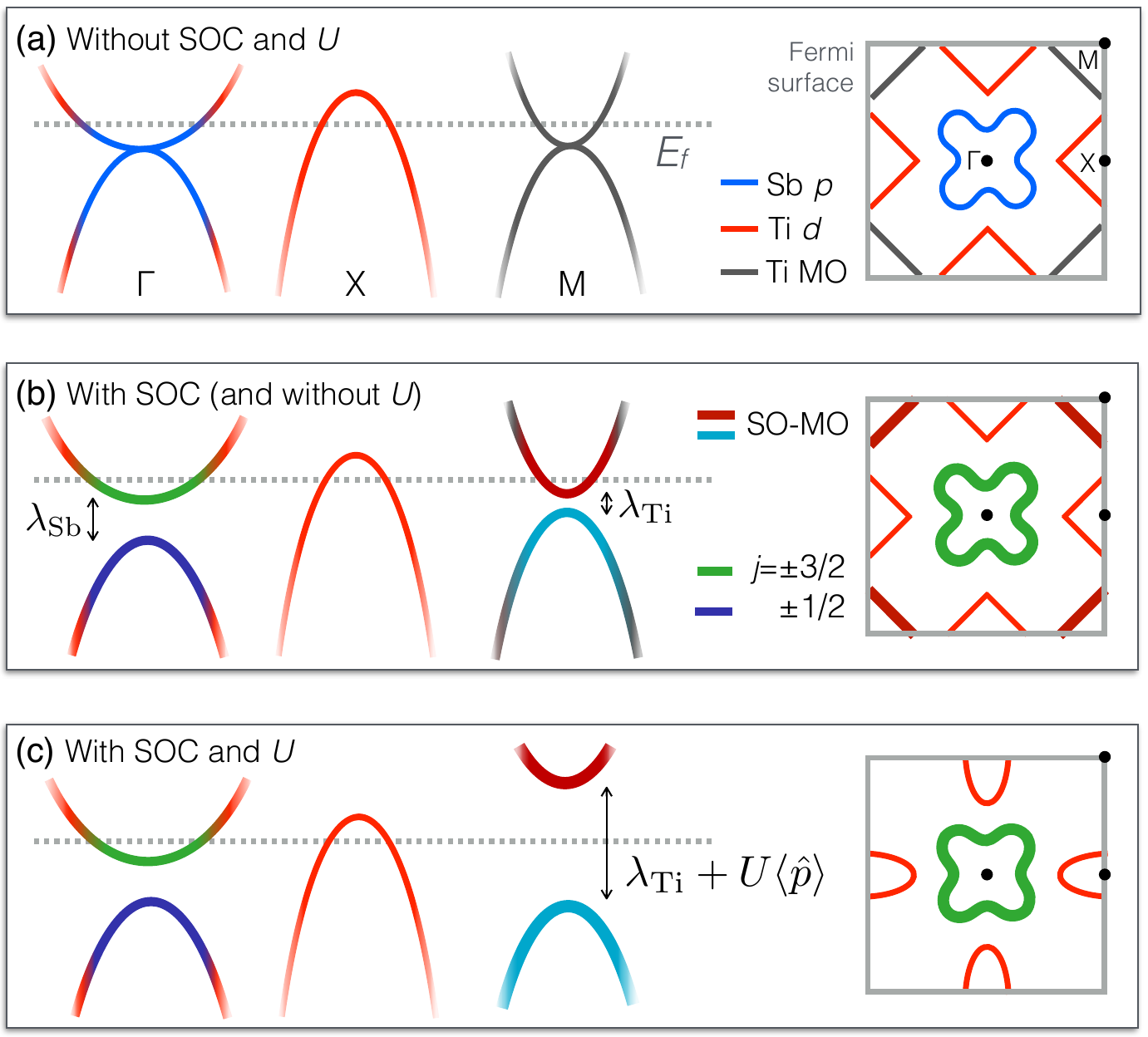}
  \caption{(Color online) Schematic figures illustrating evolution of
  electronic structures in NTSO upon inclusion of spin-orbit coupling (SOC) and on-site Coulomb 
  interaction $U$. (a) and (b) show schematic band structures near three special
  $k$ points and Fermi surfaces without and with the presence of SOC. 
  As depicted in (c), the Coulomb interaction enhances
  the polarization ($\langle \hat{p} \rangle$) 
  of the spin-orbit coupled molecular orbital (SO-MO) states near M point. Line colors 
  depict the orbital character as shown in the figure, and thicker lines represent the
  states affected by inclusion of SOC.
  	}
  \label{fig:scheme}
\end{figure} 

In this work, we provide a microscopic route to MMT without spontaneous symmetry breaking
in NTSO \footnote{In this work we employed two {\it ab-initio} density functional theory 
codes: OPENMX\cite{openmx,openmx2} and Vienna {\it ab-initio} Simulation Package\cite{VASP1,VASP2}.
Note that we did not incorporate magnetism in our calculations. Detailed parameter set adopted in both 
of the calculations are in Supplementary Material.}. 
Fig. \ref{fig:scheme} illustrates our proposal;
a momentum-dependent polarization of molecular orbitals (MO), which is
induced by the cooperation of spin-orbit coupling (SOC) and the on-site Coulomb interaction
inherent in Ti 3$d$ orbital, gaps out a significant fraction of FS
while leaving the rest of FS intact, so that the system remains metallic.
As shown in Fig. \ref{fig:scheme}(a), the system has three Fermi surfaces with atomic Sb $p$, 
Ti $d$, and MO originating from Ti $d$ orbitals at $\Gamma$, X, and M points in the Brillouin 
zone, respectively.
Degeneracies at $\Gamma$ and M points are protected by crystal and SU(2) spin rotational 
symmetries, thus introduction of SOC of Sb $p$ and Ti $d$ lifts degeneracies at $\Gamma$ and M 
respectively. Due to the smallness of Ti SOC, it makes a tiny gap at M point
and the MO states evolve into spin-orbit coupled MO (SO-MO) states as shown in Fig. \ref{fig:scheme}(b).
Inclusion of the Coulomb interaction sharply enhances the SO-MO polarization, 
such that a significant gap opens at M point, reducing the FS area in $\Gamma$ and X points
to balance the charge as shown in Fig. \ref{fig:scheme}(c).
Neither lattice nor time-reversal symmetry (TRS) are 
broken in the resulting metallic state.
Due to the lack of spontaneously broken symmetry MMT is a first-order 
phase transition, which we will discuss later.
The spin-orbit coupled and anisotropic nature of the SO-MO states
can explain the strong anisotropies in electronic and magnetic responses 
observed in recent experiments\cite{Ozawa2004,Shi2013}.
While our theory is applied to NTSO, it can be generalized to a system with different 
orbital characters which compose FS, and undergoes a partial gapping of FS across MMT. 


\begin{figure}
  \centering
  \includegraphics[width=0.47 \textwidth]{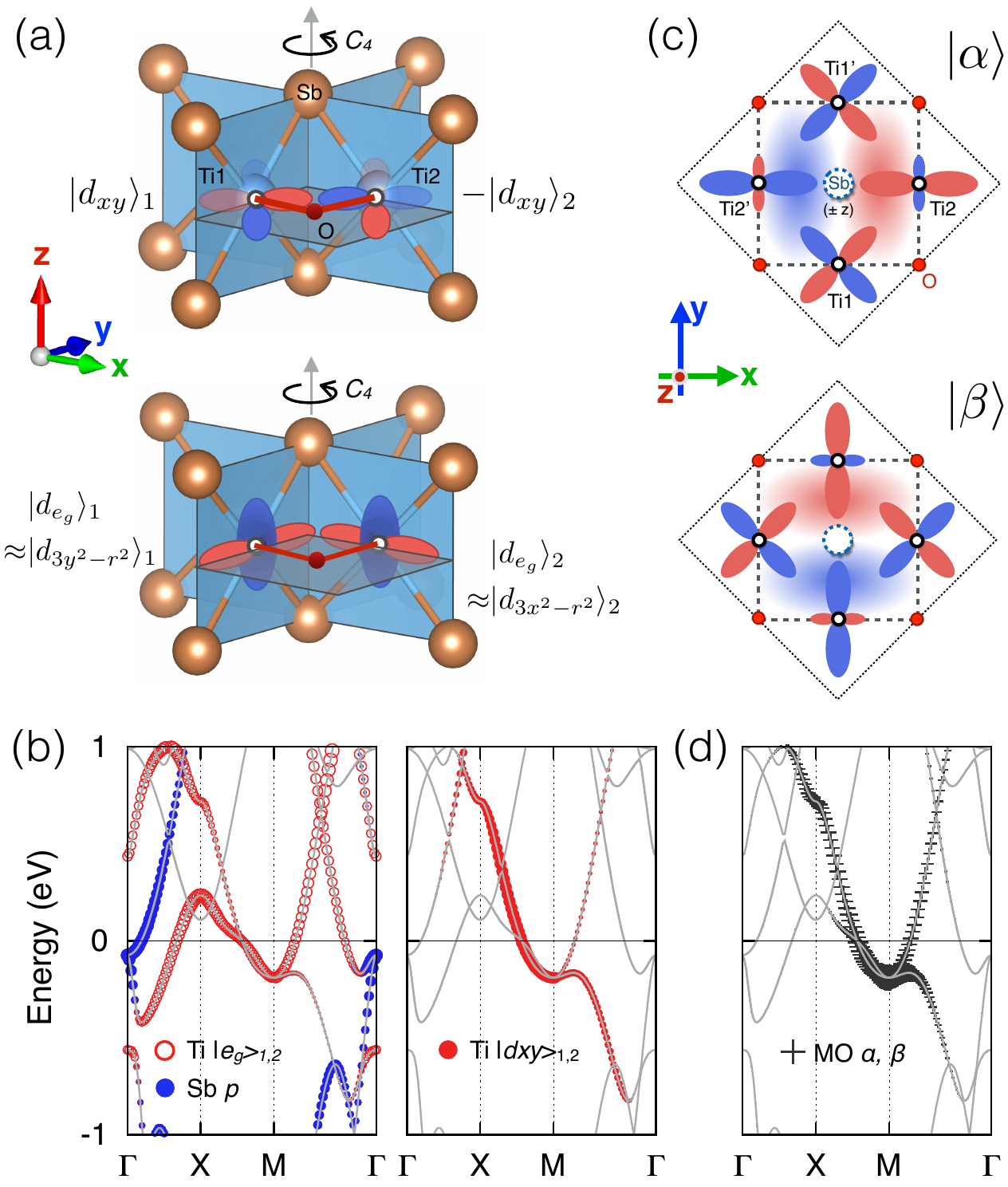}
  \caption{(Color online) (a) Crystal structure of single Ti$_2$Sb$_2$O layer
  	with dominant Ti $d$ orbitals near the Fermi level --- $\{ \vert d_{xy} \rangle_1 ,\vert d_{e_g} \rangle_1 \}$ 
    and $\{ \vert d_{xy} \rangle_2, \vert d_{e_g} \rangle_2 \}$ for Ti1 and 2, respectively ---
  	depicted in the figure. Note that Ti1 and Ti2 sites are transformed to each other
  	by the $C_4$ rotation along the $\hat{z}$ axis. 
  	(b) Band structure of single layer NTSO without including SOC and Coulomb interaction.
  	Left panel shows orbital weights of Sb $p$ and Ti $\vert d_{e_g} \rangle_{1,2}$
  	within the Bloch states with the weight proportional to the size of corresponding
  	symbols. Right panel shows Ti $\vert d_{xy} \rangle_{1,2}$ orbital weights.
  	(c) Schematic figures of two degenerate $\alpha$ and $\beta$ states at $M$ point
  	depicted in the $\sqrt{2} \times \sqrt{2}$ times enlarged unit cell (original and 
    enlarged unit cells represented as dashed and dotted square, respectively) as MOs.
  	Bands with the weights of the $\alpha$ and $\beta$ MO states are shown in (d).
  	}
  \label{fig:struct}
\end{figure}

\textit{Structure and ab-initio calculations} -- 
Crystal structure of NTSO consists of neighboring Ti$_2$Sb$_2$O layers with
Na ions intercalated in between. As shown in Fig. \ref{fig:struct}(a), 
each Ti$_2$Sb$_2$O layer is composed of a Ti$_2$O square lattice and
two TiSb$_2$ ribbons perpendicular to each other, with the unit cell
containing two Ti sites in a NTSO layer (Ti1 and Ti2 shown in Fig. \ref{fig:struct}). 
The space group for the
whole unit cell, which contains two NTSO layer, is $I4/mmm$ (No. 139), 
and for an isolated NTSO layer the layer group is $P4/mmm$ (No. 123). 

Previous {\it ab-initio} studies showed that, strong hybridization 
between the Ti $d$ and Sb $p$ orbitals yields dispersive bands for both
states
so that the system has multiple Fermi surfaces with $d$ and $p$ orbital character
as shown in Fig. \ref{fig:scheme}\cite{Pickett1998,Singh2012}.
Contrary to the $p$ orbital pocket at $\Gamma$, which shows 
three-dimensional shape, the $d$ orbital pockets show weak dispersion along 
the layer-normal direction and considered as quasi-two-dimensional Fermi surfaces. 
While we simplify the system by choosing an isolated NTSO layer as a unit cell,
comparison between the band structures from the full
and our two-dimensional unit cells in Supplementary Material\footnote{
	See Supplementary Material for computational detail and 
	further information on electronic structures.}
shows almost unaffected $d$ bands near the zone boundary by the layer stacking.


Fig. \ref{fig:struct}(a) shows the Ti $d$ orbitals which contribute to
the bands near the Fermi level; $\{\vert d_{xy} \rangle_1, \vert d_{e_g} \rangle_1\}$ 
and $\{\vert d_{xy} \rangle_2, \vert d_{e_g} \rangle_2\}$
for Ti1 and Ti2 respectively, where the subscripts 1 and 2 denote the Ti atoms 
to which the orbitals belong. The hybrid $\vert d_{e_g} \rangle_{1,2} $ orbitals
are defined as linear combinations of $e_g$ orbitals such that
 $\vert d_{e_g} \rangle_{1} \equiv -a\vert d_{x^2-y^2} \rangle_1 - b\vert d_{3z^2-r^2} \rangle_1$ 
and $\vert d_{e_g} \rangle_{2} \equiv +a\vert d_{x^2-y^2} \rangle_2 - b\vert d_{3z^2-r^2} \rangle_2$, 
where the real coefficients $a$ and $b$ are determined by the ligand fields. 
Note that, $\vert d_{e_g} \rangle_{1,2} $ are dominated by $\vert d_{3y^2-r^2} \rangle_1$ 
and $\vert d_{3x^2-r^2} \rangle_2$ respectively in our system, and $\vert d_{xy} \rangle_1$ 
and $\vert d_{e_g} \rangle_1$ are transformed to $-\vert d_{xy} \rangle_2$
and $\vert d_{e_g} \rangle_2$ respectively by the $C_4$ rotation depicted in the figure. 
Left panel of Fig. \ref{fig:struct}(b) shows the bands without SOC where the 
orbital weights of Sb $p$ and $\vert d_{e_g} \rangle_{1,2}$
states are represented as the size of the corresponding symbols. 
The $\vert d_{xy} \rangle_{1,2}$ weight is concentrated on the X-M line
as shown in the right panel in the figure.

Both the $\Gamma$ and M points have $D_{4h}$ point group symmetry which allows 
the presence of quadratic band touching points located just below the Fermi level, 
as can be seen in Fig. \ref{fig:struct}(b).
At $\Gamma$ point, the band touching consists of Sb atomic $p_{x}$ and $p_y$ orbitals
where degeneracy of the orbitals is compatible with the Sb site symmtery ($C_{4v}$).
On the contrary, the band touching at M point has dominant $d$ character
and described by molecular orbitals (MO). Without considering spin, 
two degenerate Bloch states at M point which we denote as 
$\vert \alpha \rangle$ and $\vert \beta \rangle$ are expressed as follows,
\begin{align}
\vert \alpha \rangle   &\approx \sum_{\bf R} e^{i{\bf k}_{\rm M}\cdot {\bf R}}
  \left( \vert d_{xy} \rangle_{{\bf R},1} + \vert d_{e_g} \rangle_{{\bf R},2} \right), \nonumber \\
\vert \beta \rangle   &\approx \sum_{\bf R} e^{i{\bf k}_{\rm M}\cdot {\bf R}}
  \left(  -\vert d_{e_g} \rangle_{{\bf R},1} - \vert {d_{xy}} \rangle_{{\bf R},2} \right), \nonumber
\end{align}
where ${\bf k}_{\rm M} = \left( \frac{\pi}{a}, \frac{\pi}{a} \right)$,
${\bf R}$ and $\{1,2\}$ are indices for Bravais lattice and Ti sublattices respectively.
Fig. \ref{fig:struct}(c) shows the schematic illustrations 
of the states, where the dashed and dotted squares depict
the primitive and the enlarged unit cell. 
Like the Sb $p_{x}$ and $p_y$ orbitals at $\Gamma$ point, they belong 
to the $E_{\rm u}$ irreducible representation, which are odd under spatial inversion
and transforms as $(C_4)^2 \vert \alpha \rangle = C_4 \vert \beta \rangle = -\vert \alpha \rangle$.
There is additional degeneracy
due to the SU(2) symmetry in the spin subspace without the presence of SOC, 
so the band touching at M point is fourfold degenerate and protected by the SU(2) 
and crystal symmetries. Unless either the spin or the crystal symmetry is lifted, 
the degeneracy at M point remain robust, and our LDA+$U$ calculation without
including SOC and keeping paramagnetic constraint confirms it 
(See Supplementary Material). 


\begin{figure}
  \centering
  \includegraphics[width=0.47 \textwidth]{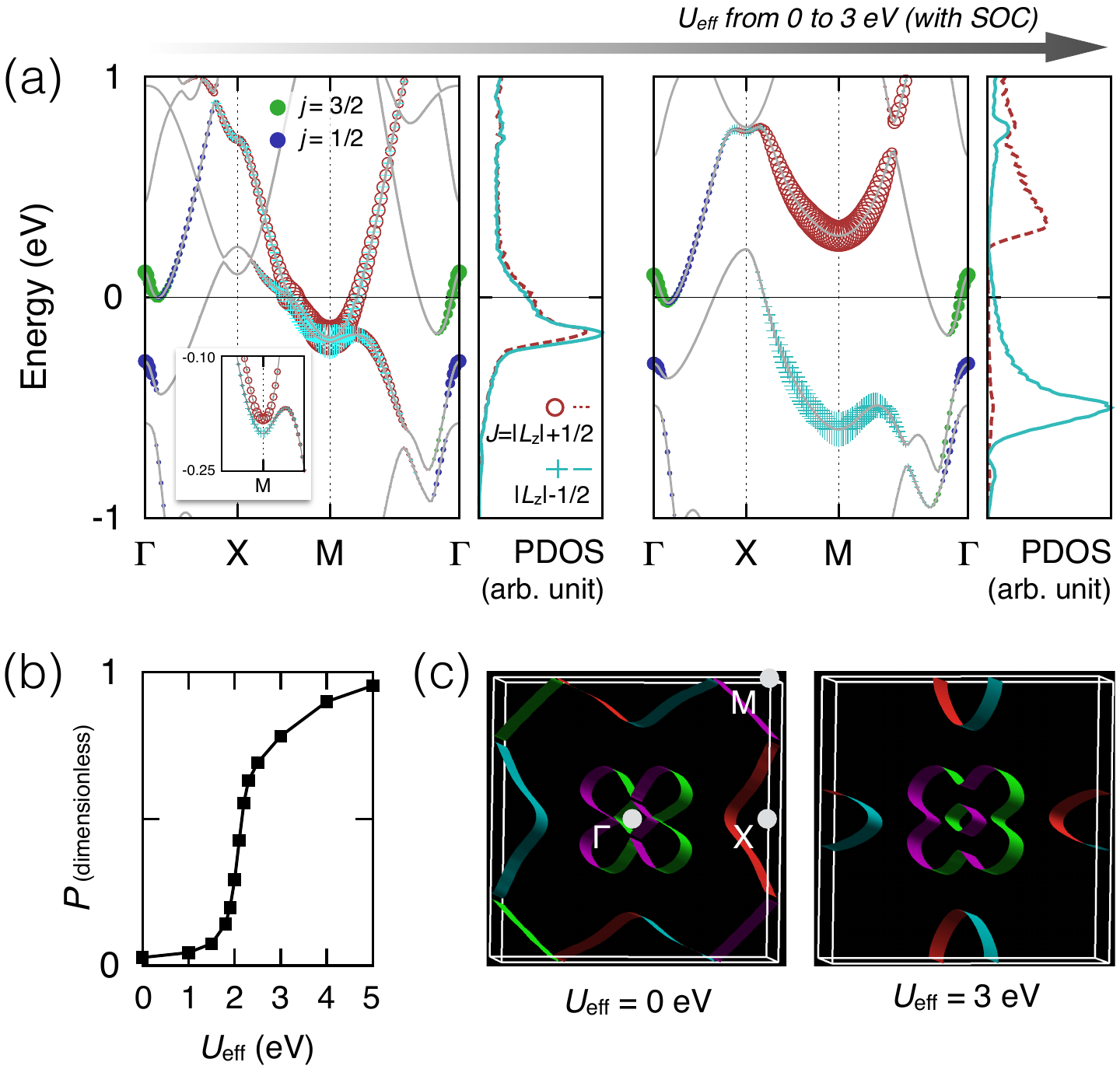}
  \caption{(Color online) (a) Band structures and PDOS projected onto the
  SO-MO states without (left panel) and with including the
  on-site Coulomb interaction ($U_{\rm eff}=3$ eV, right panel) in the
  presence of SOC. Orbital weights 
  of Sb $j= 3/2$, $1/2$, and SO-MO states are depicted as the size of 
  the symbols indicating each state. Magnified view of bands near M point
  is shown in the inset.
  (b) A plot of normalized MO polarization $P$ between the SO-MO states 
  as a function of $U_{\rm eff}$. (c) Fermi surfaces 
  of systems without (left panel) and with the Coulomb interactions (right panel).
  }
  \label{fig:SOC}
\end{figure}

\textit{Degeneracy lifting at $\Gamma$ and M points via SOC} -- 
Below we discuss the effect of SOC.
Intuitively, one expects a relatively large impact of SOC on Sb $p$ orbitals 
while negligible on Ti $d$ orbtials, as SOC in Ti is only about 20 meV.  
Indeed degenerate bands near $\Gamma$ point made of $p$ orbitals are split, 
and the gap between the two is about 0.4 eV. Each bands are characterized by 
total angular momenta
$j=1/2$ and $3/2$ made of $\vert l_z = \pm 1 \rangle = \vert p_x \pm i p_y\rangle$ 
and spin-1/2 due to ligand field spitting of layer structure.
As SOC is introduced, the SU(2) symmetry in the spin space is lifted, and the 
fourfold degeneracy at $\Gamma$ is split into $j= 1/2$ and $ 3/2$ 
doublets as shown in Fig. \ref{fig:SOC}(a).
Hereafter we denote the orbital and total angular momenta for 
the atomic and MO states as lowercase and capital letters respectively. 

Similarly, at M point, quenching of planar orbital moment components 
happens in the MO space $\{ \vert \alpha \rangle, \vert \beta \rangle \}$. 
Projecting the Ti total angular momentum operators 
$\hat{\boldsymbol{L}} \equiv \hat{\boldsymbol{l}}^{\rm Ti1} \oplus \hat{\boldsymbol{l}}^{\rm Ti1'} \oplus \hat{\boldsymbol{l}}^{\rm Ti2} \oplus \hat{\boldsymbol{l}}^{\rm Ti2'}$
onto the space yields $\hat{L}_z$ 
as the only nonvanishing component, which is diagonalized with the basis choice of 
$\vert L^\pm_z \rangle \equiv \vert \alpha \mp i \beta \rangle$
where $L^\pm_z \approx \pm \sqrt{3}$ in this system
\footnote{$L^\pm_z$ can vary from 0 to $\pm 2$
depending on the ratio of $d_{3z^2-r^2}$ and $d_{x^2-y^2}$ orbitals within
the MO states.}.
In the presence of Ti SOC, the $\vert L^\pm_z \rangle$ states are split
into two spin-orbit coupled MO (SO-MO) doublets which are characterized by
total angular momenta
$J= \left(\vert L^\pm_z \vert + \frac{1}{2} \right) \equiv J^+$ and
$\left(\vert L^\pm_z \vert - \frac{1}{2} \right) \equiv J^-$.
Note that both $J^+$ and $J^-$ form doublets with 
$J^+_z=\pm \left(\vert L^\pm_z \vert + \frac{1}{2} \right)$ and 
$J^-_z=\pm \left(\vert L^\pm_z \vert - \frac{1}{2} \right)$, respectively,
and $J^+$ doublet is higher in energy than $J^-$ by 20 meV 
at M point as shown in the inset of Fig. \ref{fig:SOC}(a).

\textit{Molecular orbital polarization enhanced by on-site Coulomb interaction} -- 
The size of the gap opening can be quantified by defining the MO polarization
operator $\hat{p}\equiv \hat{n}_{J^-} - \hat{n}_{J^+}$, where $\hat{n}_{J^\pm}$
are number operators for the $J^\pm$ subspaces.
SOC in the MO space can be rewritten by employing the MO polarization operator
as $\hat{H}_{\rm SO}=-\lambda_{\rm Ti} \hat{p}/2$. Due to the minus sign, $\hat{H}_{\rm SO}$ favors
positive MO polarization, but its magnitude is tiny because of the small $\lambda_{\rm Ti} \sim$
 20 meV as can be seen in the PDOS of Fig. \ref{fig:SOC}(a). However, once the degeneracy at M is lifted,
the size of the splitting can be further enhanced, and inclusion of on-site Coulomb interaction does the
role in this case. 

Like $\hat{H}_{\rm SO}$, the LDA+$U$ correction to the Ti on-site potential introduced by 
the Coulomb interaction can be projected onto the MO space. It can be rewritten
in terms of $\hat{p}$ as follows;\cite{Liu2008}
\begin{equation}
\hat{V}_U \equiv U_{\rm eff} \left[ \left( \frac{1}{2} - \frac{\langle \hat{n} \rangle}{4} \right)
\hat{n} - \frac{\langle \hat{p} \rangle}{2}\hat{p} \right], \nonumber
\end{equation}
where $U_{\rm eff}\equiv U-J$ is the effective Coulomb interaction parameter,
$\hat{n}\equiv \hat{n}_{J^+} + \hat{n}_{J^-}$ is the total number operator, 
and expectation values are obtained by integrating over the Brillouin zone
(detailed derivation is in Supplementary material). 
Combining it with $\hat{H}_{\rm SO}$, apart from the trivial constant term, yields
\begin{equation}
\hat{H}_{\rm SO} + \hat{V}_U =  
	-\frac{\lambda_{\rm Ti} + U_{\rm eff}\langle \hat{p} \rangle}{2} 
	\hat{p}, \nonumber
\end{equation}
so that the MO polarization $\langle \hat{p} \rangle$ initiated by SOC  
can be further increased under the presence of $U_{\rm eff}$.
Right panel in Fig. \ref{fig:SOC}(a)
shows the band structure and PDOS projected onto the SO-MO states 
with the presence of $U_{\rm eff}=3$ eV. The gap at M point
is allowed due to the loss of SU(2) symmetry via SOC, and
is greatly enhanced by the inclusion of the Coulomb interaction.
A normalized MO polarization $P\equiv \langle \hat{p} \rangle / \langle \hat{n} \rangle$ 
as a function of $U_{\rm eff}$ is plotted in Fig. \ref{fig:SOC}(b). One can see an
abrupt change in $P$ near $U^c_{\rm eff} = 2.2$ eV, where the bottom of the
$\left(\vert L^\pm_z \vert + \frac{1}{2} \right)$ band crosses
the Fermi level and the electron-like Fermi pocket near M point disappears. 
Fig. \ref{fig:SOC}(c) compares the Fermi surfaces with and without including 
$U_{\rm eff}=3$ eV. To compensate the removal of the M point electron pocket, the size of 
hole-like pocket near X point also decreases, resulting the half-reduction of the
Fermi surface area. Such reduction of Fermi surface area, or equivalently
the reduction of DOS at the Fermi level, is observed in several magnetic susceptibility
and resistivity measurements\cite{Ozawa2000,Liu2009,Shi2013}. 

We also comment on the X pocket, which is less affected by the inclusion of $U_{\rm eff}$
than the M pocket despite its dominant $\vert d_{e_g} \rangle_{1,2}$ character. Since the 
X pocket coexists with the $p$-originated $\Gamma$ pocket, the size of X pocket is determined 
by the on-site energy of $\vert d_{e_g} \rangle_{1,2}$ orbitals relative to that of 
Sb $p$. The $\hat{V}_U$ term, however, behaves as an effective SOC for the SO-MO states, 
and its contribution to the $\vert d_{e_g} \rangle_{1,2}$ on-site energy is not significant 
so that the X pocket remains even after inclusion of the Coulomb interaction. This is confirmed by examining 
the change of the on-site energies from the Wannier orbital calculations for the Ti $d$ states. 
Note also that, the Sb $j= 3/2$ and $1/2$ states near $\Gamma$ are well polarized
due to the large $\lambda_{\rm Sb}$, so they are less affected by the inclusion of $U_{\rm eff}$.



\textit{Discussion and Conclusion} -- 
The notion of orbital polarization in transition metal compounds, which is induced by 
the Coulomb interaction in degenerate $d$ orbitals, usually accompanies symmetry 
lowering by the orbital-lattice coupling such as Jahn-Teller effect\cite{OO0,OO1,OO2,OO3,Khaliullin_PTPS}. 
In metallic systems, an orbital polarization also occurs\cite{OO4}, but does not accompany 
a MMT for a fixed charge filling.
The SO-MO polarization in our work is distinguished from previous
studies since it accompanies a phase transition without spontaneous symmetry reduction. 

For experimental validation of our proposal, one can take advantage of the anisotropy in the 
SO-MO states. Since the total angular momentum of the SO-MO states are fixed to be perpendicular 
to the Ti$_2$O plane, {\it i.e.} the absence of the in-plane angular momentum component, 
one direct consquence of the SO-MO formation is an anisotropic response to the external
magnetic fields. A recent report on the magnetic susceptibility of NTSO shows such
anisotropic behavior, where the $H // z$ susceptibility data shows drastic enhancement 
compared to the $H // xy$ result below the transition temperature\cite{Shi2013}. Such behavior 
is hard to understand in the conventional charge-density wave picture, but is consistent with
our SO-MO polarization scenario; due to the absence of the $\hat{L}_{x,y}$ components in the SO-MO space, 
the states should be more susceptible to $H // z$ compared to the in-plane fields. 

Measuring the branching ratio in Ti $L_2$ and $L_3$ edge x-ray absorption spectroscopy (XAS)
below $T_c$ can be another way to experimentally validate the 
MO polarization scenario\cite{Laan_BR}. Note that, the unoccupied SO-MO state has total angular 
momentum of $j^+ =\left( \vert l^\pm_z \vert + \frac{1}{2} \right)$, which is close to $5/2$,
for each Ti site. As a result, 
$L_3$ edge XAS is expected to show higher intensity more than twice that of $L_2$ edge, 
so that the ratio between the $L_3$ and $L_2$ channels should be larger then the 
statistical ratio 2:1. Also, phase-sensitive tools such as resonant x-ray scattering
may probe the presence of SO-MO states, which revealed the presence of the spin-orbit-entangled
$j_{\rm eff}=1/2$ states in several iridate compounds with strong SOC\cite{kim2009phase,JWKim}. 

Finally, we would like to mention the nature of the phase transition at
$T_c$. Specific heat versus temperature data indicate significant loss of 
entropy at the transition. Previously this was considered as a signature of
continuous phase transition by the density wave formation, but this feature
can be explained within our SO-MO polarization scenario which results in 
entropy change by the reduction of the Fermi surface area. Since this process
does not spontaneously break any symmetries in NTSO, it should be a first-order 
transition. We found a metastable phase 
with $P \sim 0$ in the regime $2.3 \leq U_{\rm eff} \leq 4$ eV 
and its presence is robust independent of parameters and code choices
(see Supplementary Material for details on the metastable state). 
Remarkably, hystresis behavior in resistivity with unknown origin 
was reported previously\cite{Liu2009},
indicating a possible first-order phase transition.

In summary, we propose an alternative scenario for the MMT in NTSO based on the
SO-MO polarization induced by SOC and the Coulomb interaction.
Our picture can be generalized to other systems with FS with mixed orbital 
characters and spontaneous phase transition with partial gapping out of FS without
a symmetry breaking. 
Also, our picture calls attention to the role of SOC in 3$d$ transition metal compounds, 
which was considered insignificant in understanding the physics of such systems.
Indeed, there are several reports about the role of the ostensibly small SOC,
which cooperate with the Coulomb interaction and becomes a crucial element 
for the underlying physics\cite{KWLee1,KWLee2}. Further studies on the role of SOC in previously known 
transition metal systems with high crystal symmetries can be an interesting future 
subject in this regard.

\textit{Acknowledgement} -- HYK thanks the Kavli Institute for Theoretical Physics
at Santa Barbara for hospitality. This work was supported by the NSERC of
Canada and the center for Quantum Materials at the University of
Toronto.  Computations were mainly performed on the GPC supercomputer
at the SciNet HPC Consortium. SciNet is funded by: the Canada
Foundation for Innovation under the auspices of Compute Canada; the
Government of Ontario; Ontario Research Fund - Research Excellence;
and the University of Toronto.

\bibliography{NTSO}

\end{document}